\documentstyle[12pt]{article}
\input epsf

\def\1034{$10^{34}$~cm$^{-2}$~s$^{-1}$}

\begin{document}
\rightline{Princeton HEP 95-2}
\rightline{BELLE Note\# 62}
\rightline{March 31, 1995}

\bigskip
\vskip 1.25truein
\begin{center}
{\LARGE First Results from the BELLE \\ \bigskip DIRC Prototype }\\
\bigskip
\bigskip
{\large C.~Lu, D.~Marlow$^\dagger$, C.~Mindas, E.~Prebys,
W.~Sands, \& R.~Wixted}

\bigskip
\bigskip

{\large \sl Joseph Henry Laboratories, Princeton University, }

\medskip { \large \sl Princeton, NJ 08544 }

\end{center}
\vskip .75truein
\begin{abstract}
 The DIRC (Detection of Internally Reflected Cerenkov light)
is a new type of ring imaging Cerenkov detector, which detects
images from Cerenkov light produced in precisely machined quartz bars.
The Cerenkov images are transported along several meters of bar to the
edge of the detector where they are proximity focussed unto an array
of conventional photomultiplier tubes.  Results from a prototype device
comprising a $2 \times 4 \times 240$~cm$^3$ quartz bar read by an array of
480 PMT's are presented.  Sample images, which are the first observed in this
type of detector, are shown.  Measurements of the light yield
(approximately 20 photoelectrons per image) and the angular resolution are
in good agreement with Monte Carlo predictions.
\end{abstract}
\bigskip
\bigskip
\centerline{$^\dagger$ Corresponding author:  E-mail {\tt
MARLOW@PUPHEP.PRINCETON.EDU} }
\vfill\eject

\section{Introduction}

   First proposed by Ratcliff and collaborators at
SLAC\cite{RATCLIFF,RATCLIFF2}, the DIRC
({\bf D}etection of {\bf I}nternally {\bf R}eflected {\bf \v{C}}erenkov
light) technique shows great promise as a particle
ID system for $e^+e^-$ B-Factory detectors.  In particular, it promises
$\ge 4\sigma$ $\pi/K$ separation over the momentum range of interest,
occupies only $5-10$~cm of radial space inside of the calorimeter,
and places a minimal amount of material ($15\% - 20\%$ r.l.) in front
of the calorimeter.  The DIRC has been considered for the PID systems
of both the BaBar detector\cite{BABARTDR} at PEP-II and the BELLE
detector\cite{BN32,BELLETDR} at KEK.

\begin{figure}[htp]
\centerline{\epsfysize 3.75 truein
\epsfbox{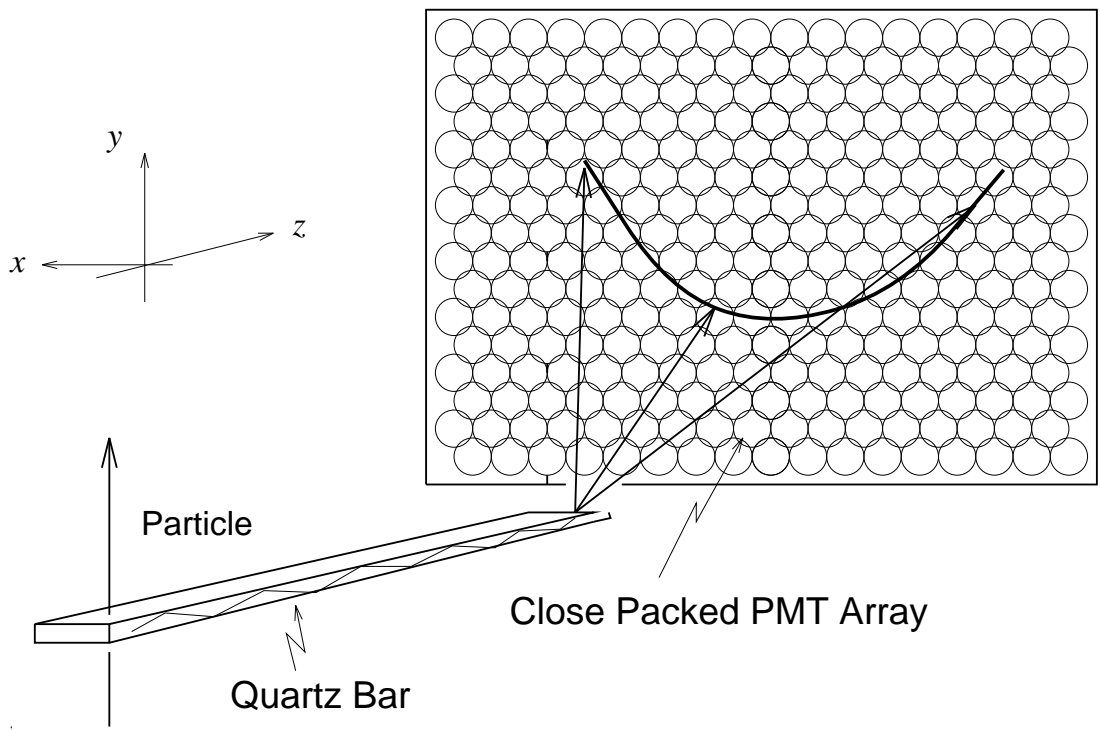}}
\caption{\label{DIRCPRIN} Principle of the DIRC.  \v{C}erenkov light
from a particle crossing the bar propagates along the bar by total internal
reflection to the end, where it emerges and is imaged onto a close-packed
array of photomultipliers.}
\end{figure}

   The DIRC concept has been described in detail in
references~\cite{RATCLIFF} \& \cite{RATCLIFF2} and will be only briefly
outlined here.  The basic idea, shown in figure~\ref{DIRCPRIN},
is that a \v{C}erenkov cone produced in
a precisely machined rectangular quartz bar (typical dimensions
$\sim 2 \times 4 \times 500$~cm$^3$) will propagate by total internal
reflection along the length bar to its end where it can be imaged by
proximity focussing onto an array of conventional photomultipliers
(more sophisticated focussing schemes are also
possible, see reference~\cite{KAMAE}).  If the surfaces of the bar
are smooth and highly rectangular, the shape of the image can be readily
calculated and provides direct information on the opening angle of the
\v{C}erenkov cone, and hence the particle's velocity.

   In this note we summarize the experimental results obtained by
the BELLE particle ID group.  Measurements of bar quality, light
yield, attenuation length, and single-photon resolution obtained
from actual DIRC images of cosmic-ray muons obtained using BELLE's
480-PMT air-standoff prototype are presented.

\section{Bar Quality}

Although the dimensional tolerances and surface quality requirements are
quite severe, it has been demonstrated that they are within the
capability of the optics industry.  In particular, the Zygo corporation
of Middlefield, Connecticut\cite{DOSS}, has produced several
120-cm-long bars of excellent quality that meet or exceed the DIRC's
requirements.

\begin{figure}[htp]
\centerline{\epsfysize 6.5 truein
\epsfbox{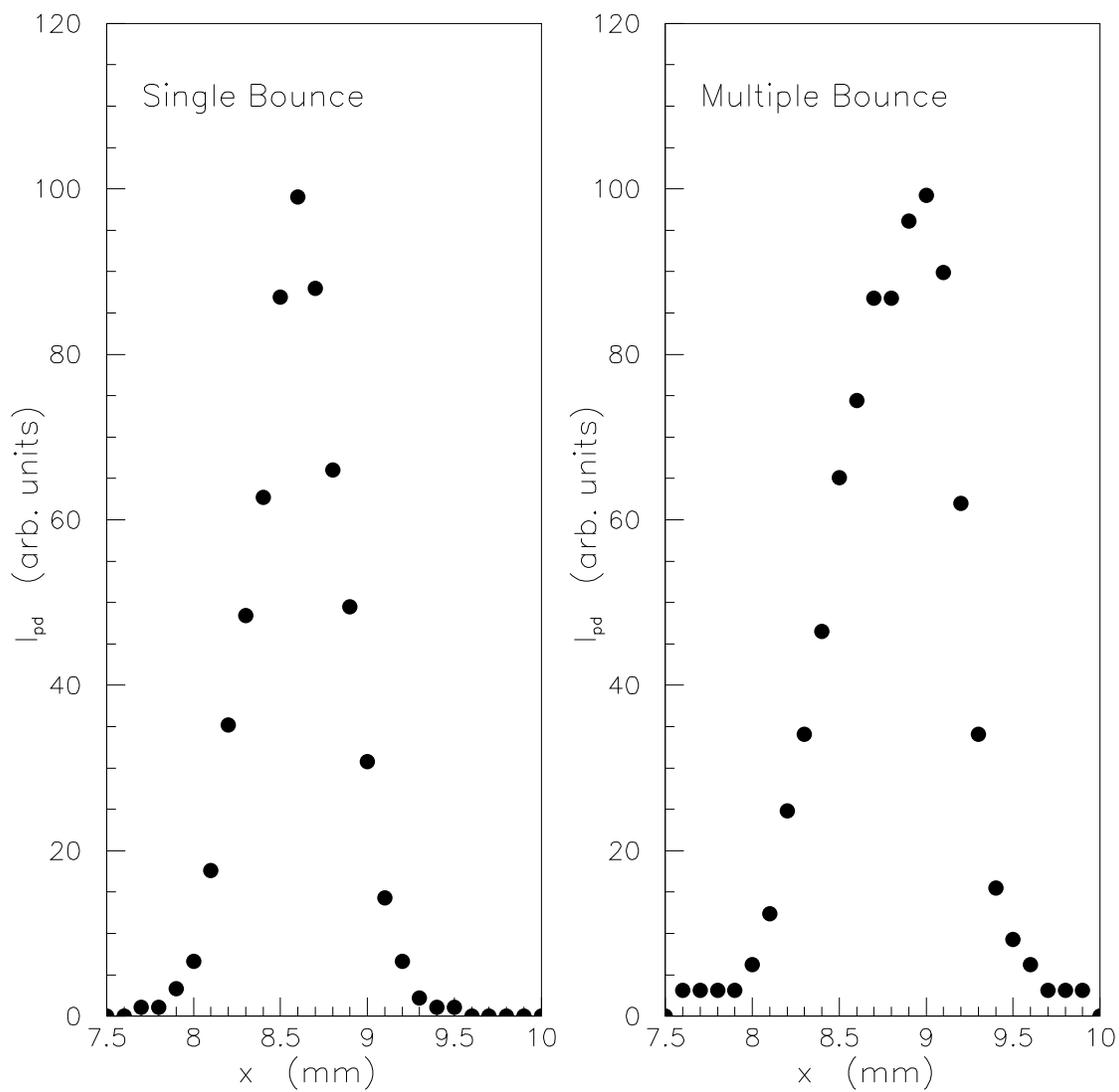}}
\caption{\label{SCAN} Beam profile of \v{C}erenkov light.  Left:
beam profile after a single reflection.  Right: beam profile after
approximately 100 reflections.  }
\end{figure}

One simple demonstration of bar quality was made using the 442~nm line from
a 10-mW He-Cd laser.  The profile of the beam was scanned by placing a
100-$\mu$m pinhole aperture in front of a PIN diode photodetector
mounted on a micropositioner.   Figure~\ref{SCAN} shows the profile of
the beam before and after a return trip along a 120-cm-long bar.  The
entrance angle of the laser was chosen such that the beam underwent
approximately 100 internal side-wall bounces before detection.   The
resulting broadening of the beam is minimal, although there is a
suggestion of a double peak, most likely the result of a corner reflection
of the laser.  However, the beam remains more than an order
of magnitude smaller than the relevant size scale of a practical detector,
which is set by the $30$-to-$40$-mm diameter of the phototubes.
Additional details on laser-based bar-quality measurements can be found
in reference~\cite{LASER}.

\section{DIRC Prototype}

\begin{figure}[htp]
\centerline{\epsfysize 5.5 truein
\epsfbox{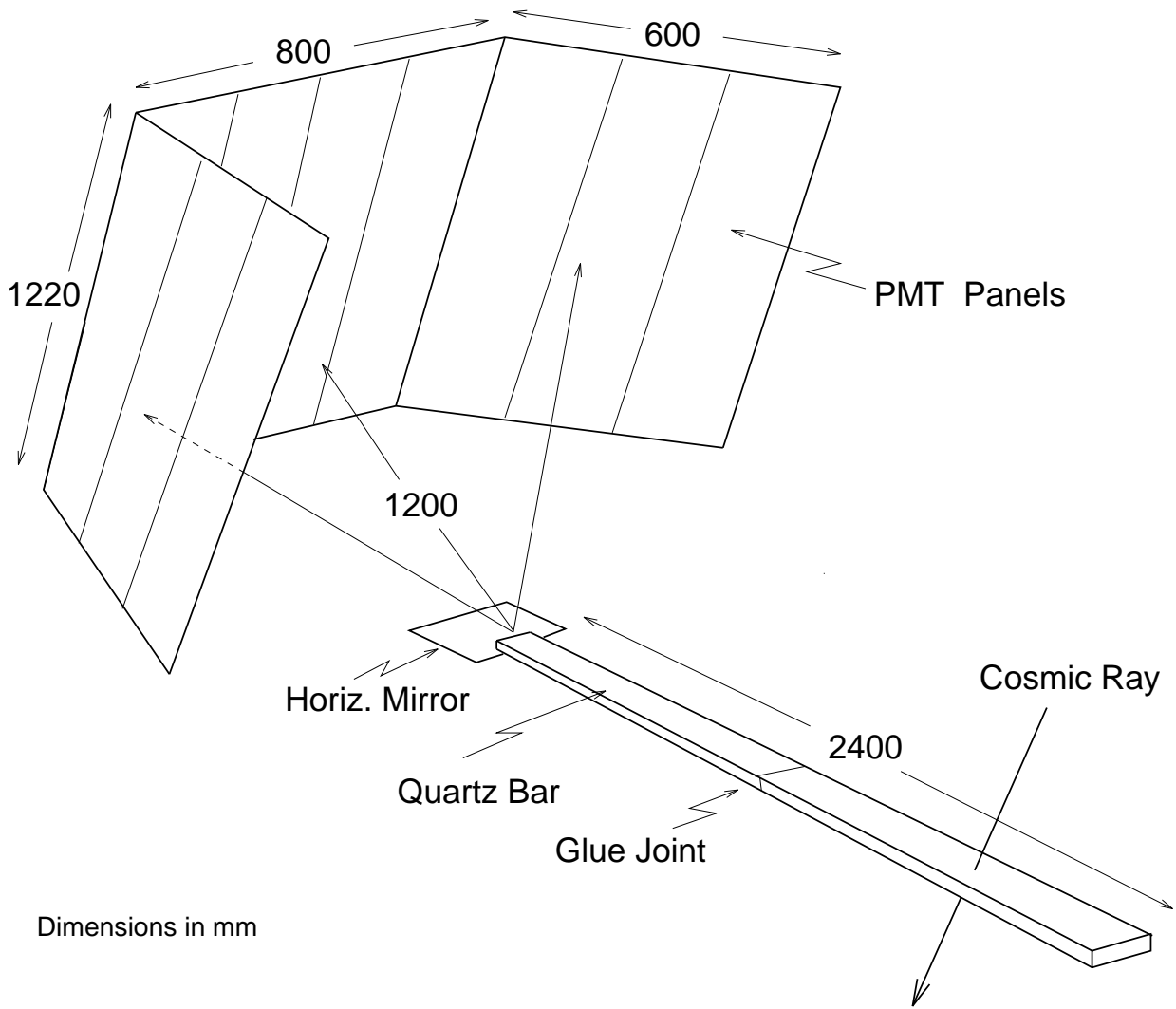}}
\caption{\label{PERSPECT} Perspective drawing of the BELLE DIRC prototype.
The PMT's (not shown) are mounted on printed circuit boards that
are attached to the vertical members of the PMT panels.  The scintillator
telescope used to select the cosmic rays is not shown. }
\end{figure}

     The results to be presented here were obtained using a
480-PMT air-standoff
DIRC, which is shown schematically in figure~\ref{PERSPECT}.
Additional details can be found in reference~\cite{BN39}.
Two 120-cm-long bars were glued together using Epotek 305 epoxy to form
a single $2 \times 4 \times 240$~cm$^3$ bar.  A $15 \times 15~{\rm cm}^2$
horizontal mirror was placed at the readout end of the bar to deflect
downward heading light emerging from the bar in the direction of the array.
The face of a typical phototube in the array
was situated 120~cm from the end of the bar.
Although in a production device the
standoff region (the volume between the quartz
bar and the phototube array) will be filled with water, for this test,
which was intended as a proof of principle, an air standoff was employed.
This greatly simplified the engineering of the prototype.

   The photodetector array comprised 480
38-mm-diameter 10-stage (Hamamatsu Model R580) PMT's mounted on ten
separate HV-distribution/signal-collection
printed-circuit boards.  The PMT's were sorted by gain into groups of
sixteen.   Each such group shares a common HV divider string,
substantially reducing the required number of HV channels and simplifying
the assembly.  The signals from the phototubes were readout using
GASSIPLEX\cite{SANTIARD} chips developed at CERN for pad-chamber readouts.
The GASSIPLEX chips were mounted in close proximity to the PMT's, which
were coupled using a 5:1 capacitive divider.
Although the 700~ns integration time of the GASSIPLEX chips would render
them unsuitable for a production device (the inherent speed of the PMT's
represents an important advantage for the DIRC), for cosmic-ray work, where
the rate is low, it produces a convenient
delay that eliminates the need for bulky trigger-delay cables.  Moreover,
the highly multiplexed nature of the GASSIPLEX allows us to read the
entire array with a single CAMAC ADC module.

Cosmic-ray
muons incident over an angular range $ 20^\circ < \theta_t < 30^\circ$
were selected using a three-scintillator telescope and a 40-cm-thick steel
filter.  The muons were tracked using a set of 6-mm-diameter straw
chambers operating as drift chambers in limited streamer mode.  These
were arranged so as to provide two double-layer measurements in each view,
thereby allowing measurements of $\theta_t$ and $\phi_t$ (the angles the
muon makes with the vertical in the planes parallel to
and normal to the bar's axis, respectively) and completely
determining the muons' incident trajectories.

   The data reported here were taken in three sets of runs at three
different locations along the bar.  Two of the datasets were taken
in ``forward readout'' mode---i.e. the angle of the telescope ($25^\circ$
to the vertical) was chosen so as to direct the \v{C}erenkov light in
the direction of the array.  For these two datasets the distances between the
telescope location and the readout end of the bar were 75~cm and 175~cm,
referred to as ``$z=75$~cm'' and ``$z=175$~cm,'' respectively.  For the third
dataset, a ``backward readout'' mode was employed, wherein the orientation
of the telescope was reversed (but kept at $25^\circ$
to the vertical) such that the \v{C}erenkov cone was directed
initially away from the readout end of the bar.  For this dataset, a mirror was
placed at the end of the bar opposite from the readout to redirect the
light back in the readout direction.  Since the
total distance (measured along the bar) covered by the light in this mode
was approximately 410~cm, this data is referred to as the ``$z=410$~cm''
sample.

\section{Results}

\begin{figure}[p]
\centerline{\epsfysize 7.5 truein
\epsfbox{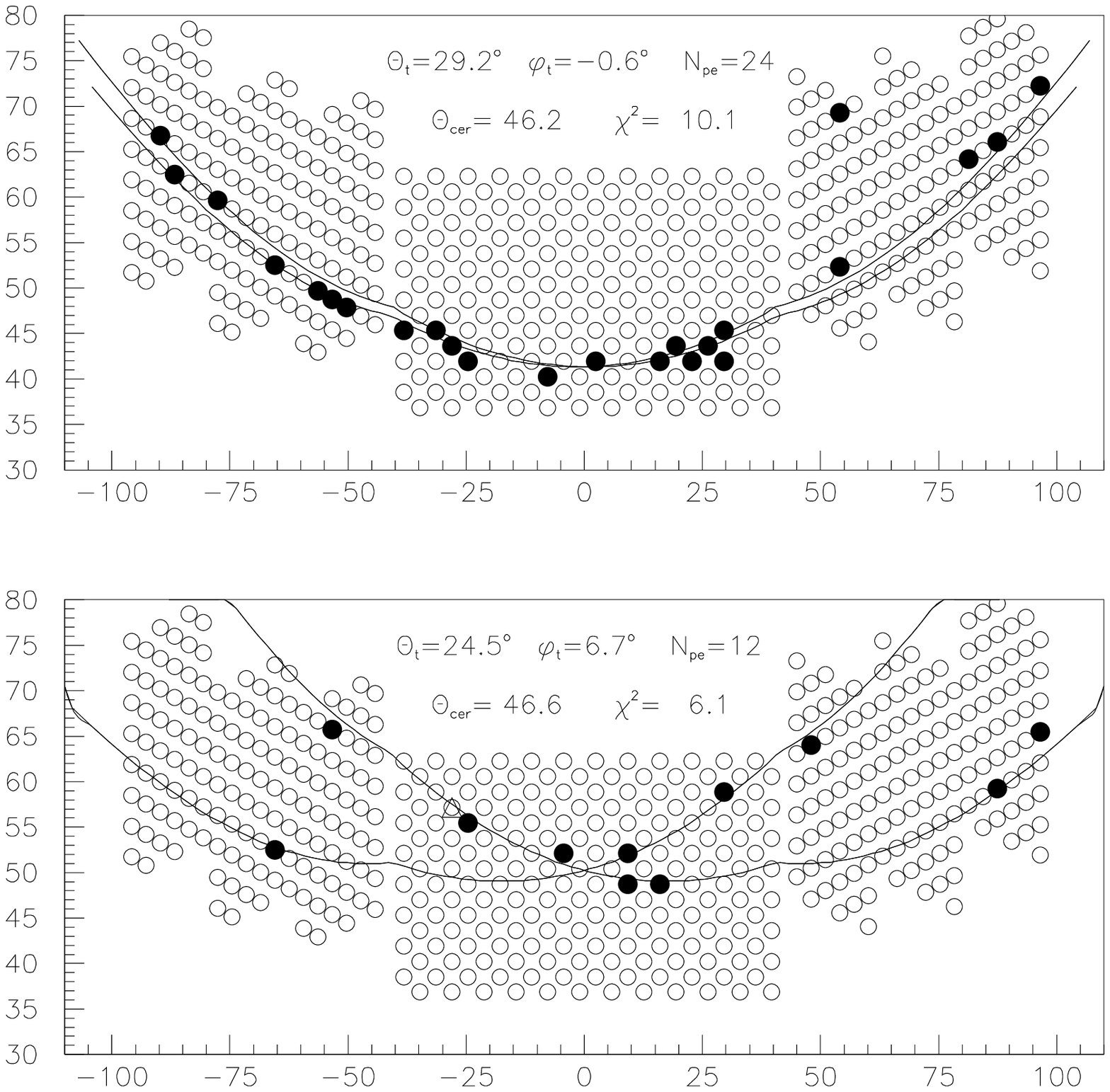}}
\caption{\label{EVENTS1} Two typical ring images.}
\end{figure}

\begin{figure}[p]
\centerline{\epsfysize 7.5 truein
\epsfbox{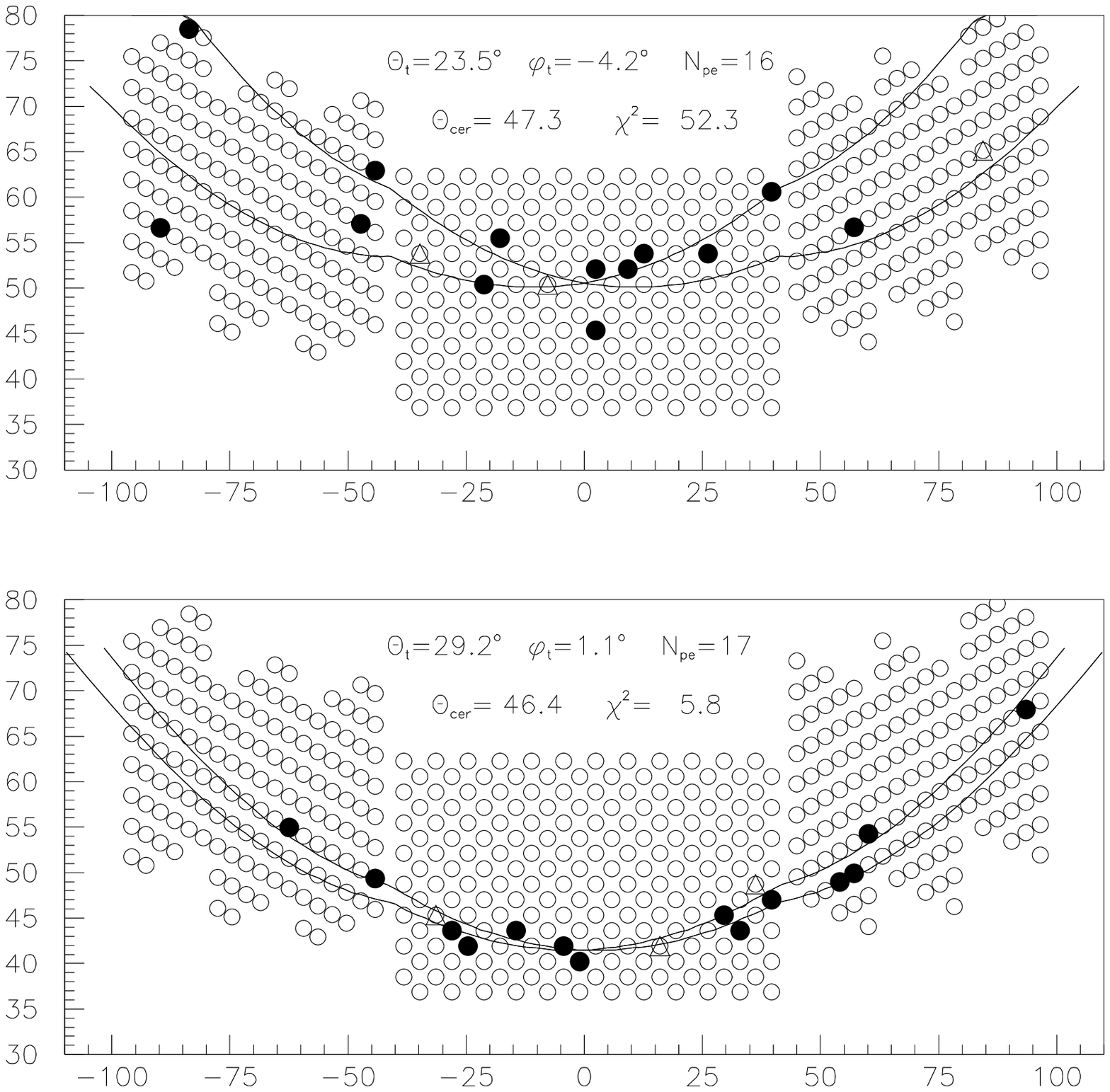}}
\caption{\label{EVENTS2} Two more typical ring images.}
\end{figure}

The excellent performance of the DIRC is quite evident at
a qualitative level from the four event displays shown in
figures~\ref{EVENTS1} and \ref{EVENTS2}.
These images were
obtained within hours of turning on the array and are quite typical.
The open circles show the positions of the phototubes if projected
onto a flat plane coincident with the center panel.  The solid circles
represent struck phototubes and the open triangles represent struck
phototubes with pulse heights that are small compared to the average
for single photoelectrons but still above threshold.

The position and shape of the curves depend on the incident directions
of the muons, which were determined by the straw chambers, and on
the \v{C}erenkov angle, which was taken as the single free parameter in
a curve fitting position.  Due to the left-right ambiguity in the
reflection of the light (i.e., the number of horizontal bounces
for a given photon can be either even or odd) there are in general
two ``solutions.''   In cases where the incident muon is perfectly
vertical, these two solutions coalesce, as is (nearly) the case in
upper panel of figure~\ref{EVENTS1} and the lower panel of
figure~\ref{EVENTS2}.

   In the sections that follow we present the results of a preliminary
analysis of the quantitative aspects of the DIRC's performance.

\subsection{Photoelectron Yield}

  We first digress to summarize some basic notions regarding photoelectron
yield calculations and some results obtained with a single PMT readout.
A simple model for the light yield is
\begin{equation}
N_{\rm p.e.}^{\rm obs} = N_0 \left<\sin^2\theta_C\right> \varepsilon_{\rm geom}
\end{equation}
where $N_{\rm p.e.}^{\rm obs}$ is the observed (or predicted) photoelectron
yield, $\left<\sin^2\theta_C\right>$ is a suitably averaged quantity
determined mainly by the refractive index of the quartz, and
$\varepsilon_{\rm geom}$ is the geometric acceptance, which can be
reliably calculated by Monte Carlo.  The least certain factor is $N_0$,
the \v{C}erenkov quality factor.  In
general $N_0$ depends on the quantum efficiency of the photocathode and on the
transmission of the PMT window material (both functions of wavelength).
Although in principle these quantities can be measured, a
more practical approach is to determine $N_0$ by using the measured
value of $N_{\rm p.e.}^{\rm obs}$ from a well understood
geometry---i.e., a bar read at one end by a PMT.

Such a measurement was carried out using a Burle 8850 ``Quantacon'' PMT
coupled to the bar with with $n=1.40$ (GE Viscosil 600) grease.  A
yield of $66 \pm 0.6~(\rm stat)$~p.e.'s, was observed for a
sample of hardened cosmic-ray
muons incident at an average angle of $\theta=30^\circ$ to the
normal.   This value was used
to deduce $N_0=121$~p.e./cm.  As a check, the measurement was repeated
with air coupling to the bar, where Monte Carlo predicts the yield should be
50\% of that for grease.  The observed yield was
$32.8 \pm 1.7~(\rm stat.)$~p.e.
to be compared with the prediction of $33.4$.    The systematic error in
the measurements is estimated to be 10\%, due mainly to the systematic
uncertainty in fitting the single photoelectron peak to determine the
phototube gain and to uncertainties in the geometry of the cosmic-ray
telescope.  These measurements are in good agreement
with previously reported results\cite{IEEE} and with results
recently obtained in beam tests at KEK\cite{KICHIMI}.

\subsection{DIRC Prototype Yield Measurements}
\label{YYY}
\begin{figure}[p]
\centerline{\epsfysize 6.5 truein
\epsfbox{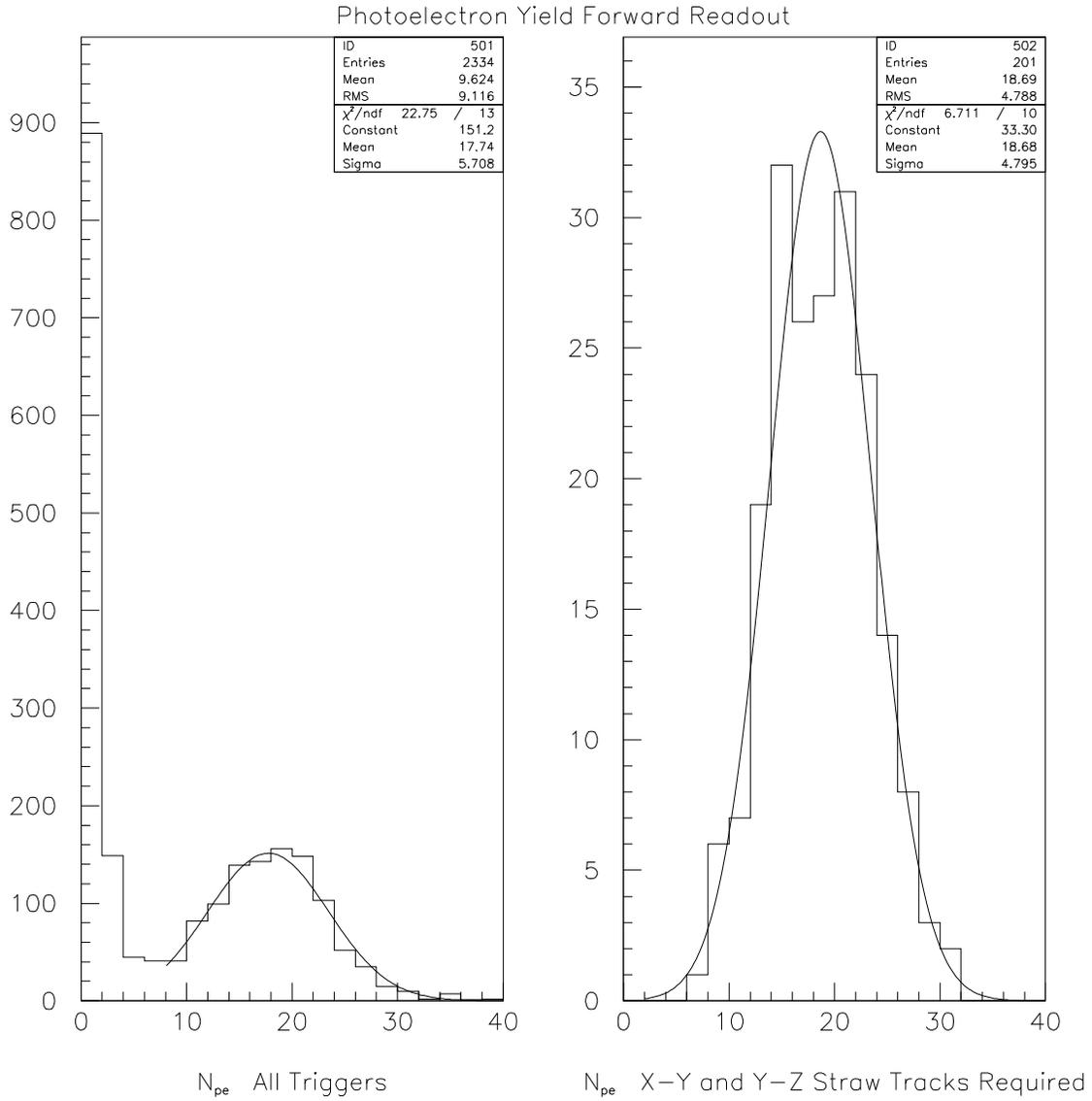}}
\caption{\label{YIELD} Photoelectron yield for the prototype array in
forward readout mode.
Left: number of struck PMT's for all triggers.  Right: number of struck
PMT's for triggers that result in straw-chamber tracks that pass
through the bar.   The data include runs from both $z=75$~cm and $z=175$~cm. }
\end{figure}

    Figure~\ref{YIELD} shows the photoelectron yield for the DIRC prototype
operating using cosmic rays incident at $25^\circ$.  The data include
runs at both $z=75$~cm and $z=175$~cm (the yields are the same within
statistics).  The yield is defined as the number of
PMT's with ADC values that are 10$\sigma$ above pedestal (pedestal
widths are typically four ADC counts on a 10-bit scale)\footnote{We estimate
that there is a 5-10\% loss of photoelectrons due to this threshold,
but we have not included it in our yield calculations.}  The observed
yield is $N_{\rm p.e.} = 18.5 \pm 0.5$.  The distribution is reasonably
gaussian.   Note that for the right-hand plot in figure~\ref{YIELD},
the only cuts are on the straw tracking chambers.  No requirements are
placed on the DIRC array itself.

   A Monte Carlo calculation of the expected yield using $N_0 = 121$~p.e./cm
has been made.  Including two approximately 4\% losses due to
Fresnel reflections
at the quartz-air and air-PMT-window interfaces, the
predicted yield is $N_{\rm p.e.} = 18.5$.   We estimate the systematic
error to be approximately 10\%.  The close agreement between
Monte Carlo and data is no-doubt fortuitous, but reassuring nonetheless.
(Also, as discussed in section~\ref{RES} below, the usable yield for
these runs is actually 3\%-6\% lower.)

\begin{figure}[p]
\centerline{\epsfysize 6.5 truein
\epsfbox{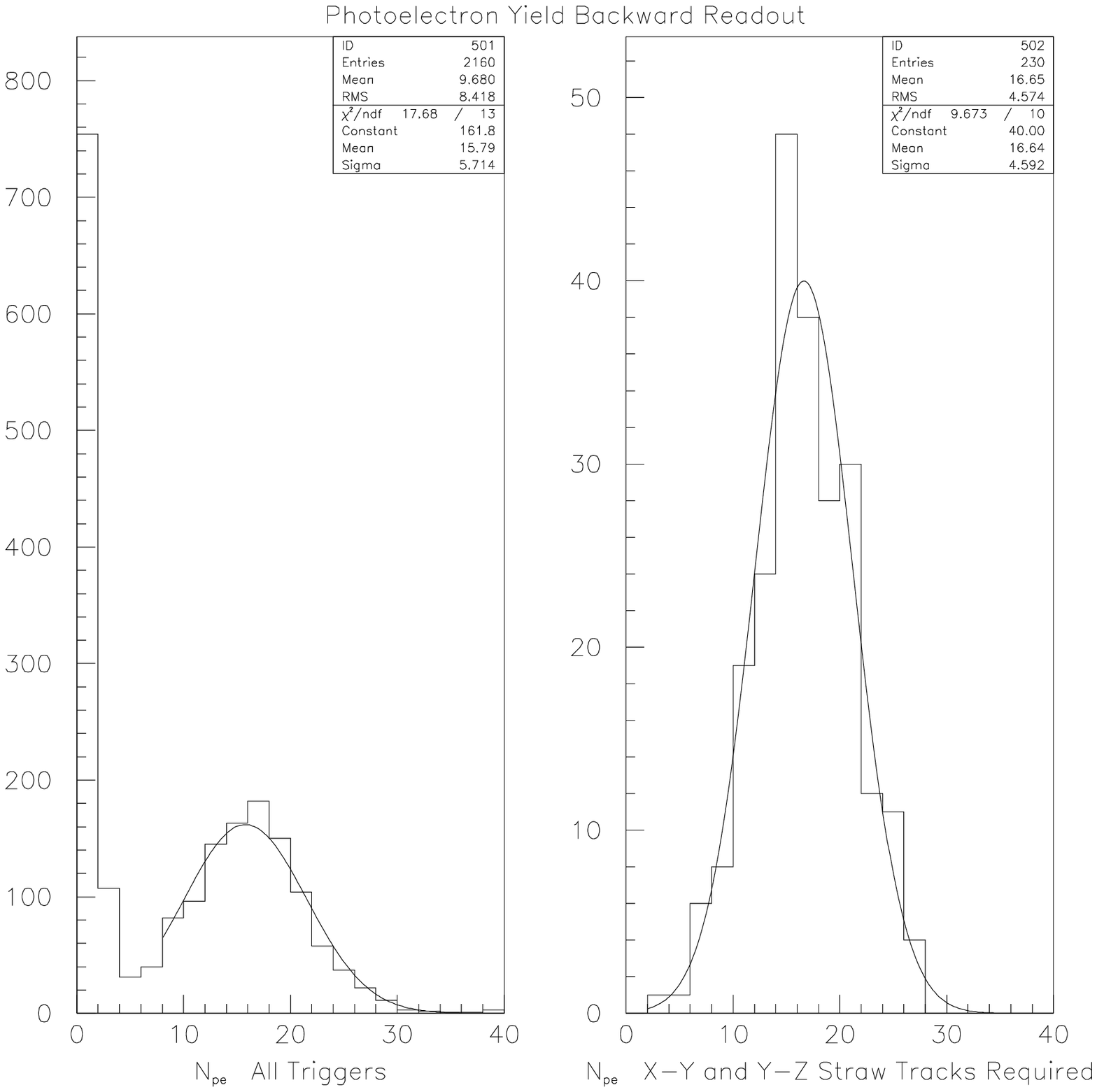}}
\caption{\label{YIELDW} Photoelectron yield for the prototype array
in backward readout mode.
Left: number of struck PMT's for all triggers.  Right: number of struck
PMT's for triggers that result in straw-chamber tracks that pass
through the bar.   The data are from the $z=410$~cm dataset. }
\end{figure}

    Figure~\ref{YIELDW} shows the photoelectron spectra for the $z=410$~cm
data.  According to the vendor's datasheets the broadband reflectivity
of the end mirror is approximately 90\% (the Fresnel reflections have
little effect  since they also direct the light back along the bar).  Taking
into account the additional path length along the bar of $\sim 285~{\rm cm}$
one expects an additional loss of about 15\% from bar attenuation\footnote{This
estimate is based on attenuation length measurements recently reported
by Kichimi {\it et al.} at KEK, who report a good fit to their data assuming
a surface reflectivity of $r=.9996 \pm .0003$}.
Relative to the forward readout data we thus
expect $N_{\rm p.e.} \simeq 0.9 \times 0.85 \times 18.5
= 14$ for the backward readout.  The observed value is
$N_{\rm p.e.}= 16.6 \pm 0.5$, somewhat higher than expected.
We note, however, that the point-to-point
systematics of these measurements are not well controlled  (for
example, the cosmic ray telescope must be completely disassembled each
time the readout position is changed)  and variations at the 10\% level
cannot be ruled out.

\subsubsection{Single Photon Resolution}
\label{RES}
   The \v{C}erenkov angle resolution (or ``ring'' resolution) is approximately
given by
\begin{equation}
\sigma_{\rm ring} \simeq { \sigma_{\gamma} \over \sqrt{N_{\rm p.e.}}}
\end{equation}
where $\sigma_{\gamma}$ is the per-photon angular resolution
and $N_{\rm p.e.}$ is the yield.  The
main contributions to $\sigma_{\gamma}$ come from chromaticity in the
radiator and the spatial resolution of the
photodetector.   The latter is driven by the transverse dimensions of
the bars, the size of the phototubes and the standoff distance.  Both
effects are included
in the Monte Carlo.  The spread in \v{C}erenkov angles for muons
passing through the bar and the steel range stack is estimated to be
$\sigma_{\rm ring}^{\rm cosmic} \sim 5$~mrad.  Since this is not negligible
compared to the expected $\sigma_{\gamma}$, we have elected to plot
a quantity called $\Delta \theta_C$, which is given by
\begin{equation}
\Delta\theta_C = \left| {d \theta_C \over ds} \right| \Delta s
\end{equation}
where $\Delta s$ is the distance of closest approach between the
measured hit position and the fitted curve and $d \theta_C/ds$, which
relates variations in $\Delta s$ to changes in the
\v{C}erenkov angle, is calculated numerically for each photon.
Note that since $\Delta \theta_C$ is in effect a fit residual, it
strictly speaking is not the same as $\sigma_\gamma$.  However,
given that there are typically 15-20 photoelectrons  per image, the
two quantities should be reasonably close in value. In any event, since
the same procedure is applied to both Monte Carlo and data it is
possible to check for consistency.

\begin{figure}[p]
\centerline{\epsfysize 7.0 truein
\epsfbox{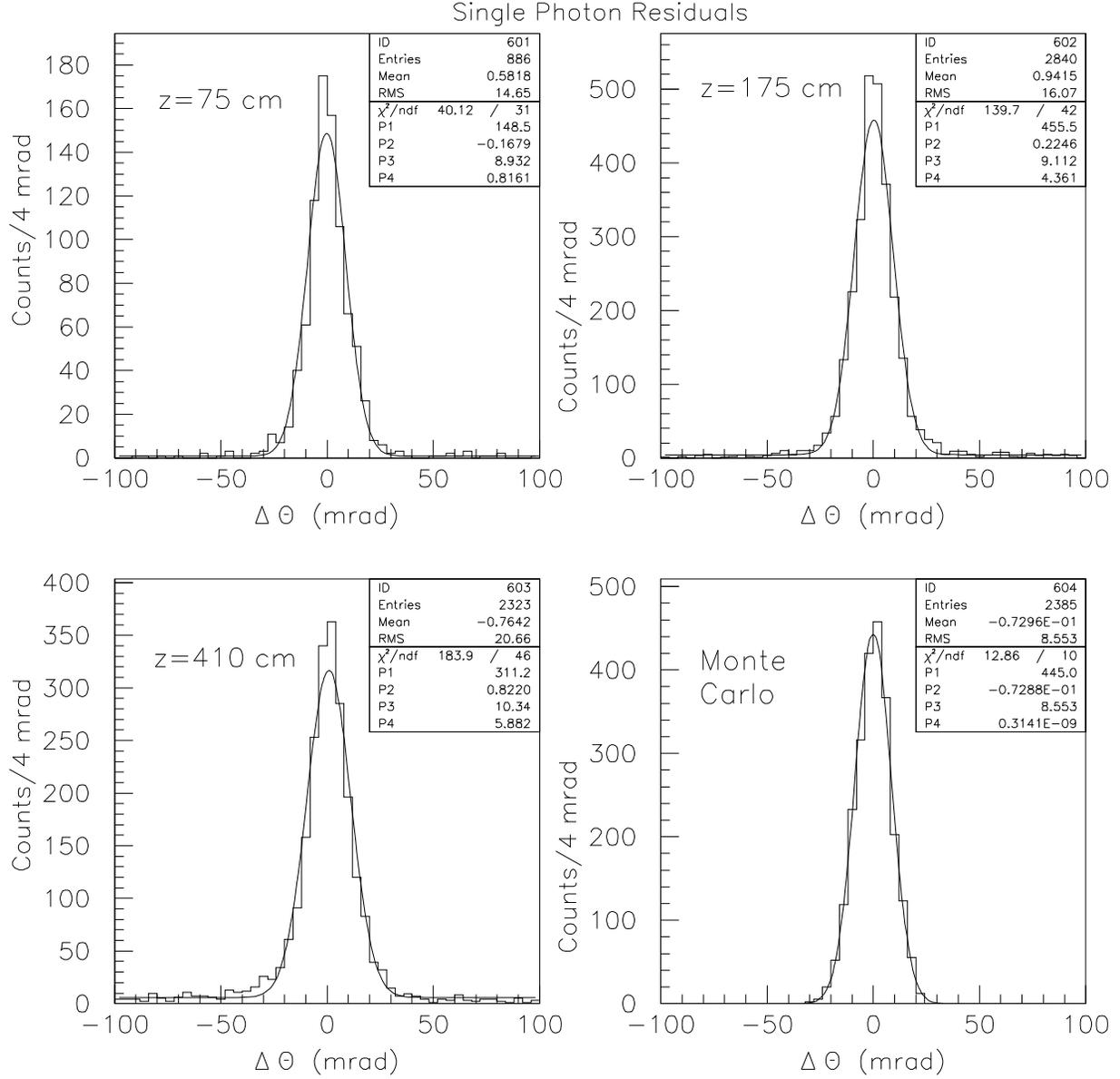}}
\caption{\label{DTHTMN3} Distribution of $\Delta \theta_C$ for the various
datasets and for Monte Carlo.  The parameters P1, P2, and P3 correspond
to the fitted amplitude, mean, and $\sigma$ of the gaussian term.  P4 is
the amplitude of the flat background.  }
\end{figure}

Figure~\ref{DTHTMN3} shows the distribution of $\Delta \theta_C$ for
the three datasets and for Monte Carlo. If a flat background term
is added to the fits, the distributions are approximately, although by no
means perfectly, described by single gaussians of $\sigma \simeq 9$-10~mrad
Not surprisingly, the Monte Carlo distributions are more nearly
gaussian with $\sigma=8.6$~mrad---reasonable agreement given the maturity of
the analysis.

  There are some differences between data and Monte Carlo.  First, there
is a small, but statistically significant, difference in width between
the width of
the forward readout and backward readout
datasets.  The origin of this effect is under
investigation, but is not currently known.  The possibilities include
dimensional imperfections in the quartz bars, performance variations in
the cosmic-ray telescope (the angular resolution of the tracking chambers,
for example, has an influence on the apparent photon resolution), and
alignment of the forward mirror.

There is also a small uniform background whose magnitude appears to grow
with distance.\footnote{the ranges of the histograms in figure~\ref{DTHTMN3}
correspond roughly to the size of the array, although there are a few events
just outside of the displayed range.}   If one defines $\eta_{\rm bkgr}$
as
\begin{equation}
\eta_{\rm bkgr} \equiv
1 - { \#~{\rm photons\ with}\ (-35~{\rm mrad} < \Delta\theta_C <
35~{\rm mrad}) \over {\rm total\ \#~photons} }
\end{equation}
then $\eta_{\rm bkgr}= 3.6 \pm 0.6\%,\ \ ~4.9 \pm 0.4\%$, \ and \
$8.4 \pm 0.6\%$ (statistical errors) for the
75-, 175-, and 410-cm
data, respectively.  These hits are not predicted by the Monte
Carlo, nor can they be attributed to solely to random backgrounds.   From
an analysis of pulser triggers the random background rate per event is
0.37 photons, which if the only source of background would result in
$\eta_{\rm bkgr} \simeq  1.5\%$.     Possible origins of the excess
include small-angle scattering in the bulk quartz,  surface imperfections,
and scattering at the glue joint (this would not account for the
$z=75~$cm data).    We note that since these photons do not contribute
to the ring resolution, they should be subtracted from the yield numbers
obtained in section~\ref{YYY} above.

\section{Summary and Conclusion}

    Preliminary test results from the BELLE DIRC prototype have been
presented.
The performance of the prototype is reasonably well described by
the Monte Carlo, although there are some as-of-yet unresolved differences,
which are currently under investigation.
On balance, the performance of the device is remarkably good considering
the early stage of development of  DIRC technology.

\section{Acknowledgements}

   We would like to thank F.~Doss, T.~Kamae, H.~Kichimi, K.~McDonald,
S.K.~Kim, Y.~Sakai, F.~Shoemaker, S.~Uno and Y.~Watanabe, for useful
discussions
and B.~Ratcliff of SLAC, who was very generous in sharing his expertise and
in lending us a quartz bar for our initial laser tests.  K.~Abe, S.~Olsen,
M.~Selen, and F.~Takasaki provided helpful advice and encouragement as well
as material support in the form of phototubes and quartz bars.
V.~Polychronakis of BNL was kind enough to lend us the ADC readout
module.  Finally, we are
also very grateful to C.~Bopp, S.~Chidzik, W.~Groom, R.~Klemmer, S.~Morreale,
A.~Nelson, and R.~Rabberman, for their excellent technical work.


\begin{thebibliography}{99}


\bibitem{RATCLIFF} B.~Ratcliff, $Ba{\overline B}ar$
Collaboration Note~\#92 (Dec. 1992);
P.~Coyle {\it et al.}, SLAC Report SLAC-PUB-6371, (1993).

\bibitem{RATCLIFF2} B.~Ratcliff, ``The DIRC Counter: A New Type of Particle
Identification Device for B Factories,'' Proceedings of the International
Workshop on B-Factories: Accelerators and Experiments, pg. 331, KEK Proceedings
93-7, (November 1992).

\bibitem{BABARTDR} BaBar Detector Technical Design Report, March 1995.

\bibitem{BN32} ``The DIRC Option for BELLE,'', D.~Marlow, BELLE Note~32,
August 1994.

\bibitem{BELLETDR} BELLE Detector Technical Design Report, KEK Report,
January, 1995.

\bibitem{KAMAE} ``Focussing DIRC as the Particle Identifier for
the Forward End-Cap and Barrel Regions,'' T.~Kamae {\it et al.},
BELLE Note 49, January 1995.

\bibitem{DOSS} F.~Doss, Zygo Corporation, Middlefield, Connecticut.


\bibitem{LASER} ``Optical Tests of a Quartz DIRC Radiator Bar,'', C.~Lu
and D.~Marlow, BELLE Note (in preparation).

\bibitem{BN39} ``BELLE DIRC Phase I Prototype Notes,'' S.~Kanda {\it et al.},
BELLE Note 39, November 1994.

\bibitem{SANTIARD} The GASSIPLEX is a redesign of the AMPLEX chip by
J.C.-Santiard of CERN.  The original AMPLEX is described in
E. Beuville {\it et al.}, Nucl. Inst. and Meth. A288 (1990) 157.

\bibitem{IEEE} D.~Aston {\it et al.}, talk by H.~Kawahara at the
IEEE 1994 Nucl. Sci. Symp., Norfolk VA, October 1994 (SLAC-PUB-6731).

\bibitem{KICHIMI} H.~Kichimi, private communication.

\bibitem{BABARLOI} PEP-II Collaboration Letter of Intent, June 1994.

\end{thebibliography}
\end{document}